\documentclass[aps,nofootinbib,superscriptaddress,twocolumn,prd,10pt]{revtex4-1}

\usepackage{graphicx}
\usepackage{amsmath,amssymb}
\usepackage{hyperref}
\usepackage{braket}
\usepackage{booktabs}
\usepackage{subfigure}
\usepackage{float}
\usepackage[dvipsnames]{xcolor}
\usepackage{mathrsfs}
\usepackage{comment}
\usepackage[normalem]{ulem}

\hypersetup{
    colorlinks=true,
    linkcolor=red,
    citecolor=blue,
}

\newcommand{\wn}{w_{\rm f}}
\newcommand{\ADE}{{\rm ADE}}
\newcommand{\fADE}{{\rm ADE}}

\begin{document}
\title{Testing $H_0$ in Acoustic Dark Energy Models with Planck and ACT Polarization}

\author{Meng-Xiang Lin}
\affiliation{Kavli Institute for Cosmological Physics, Department of Astronomy \& Astrophysics, Enrico Fermi Institute, The University of Chicago, Chicago, IL 60637, USA}

\author{Wayne Hu}
\affiliation{Kavli Institute for Cosmological Physics, Department of Astronomy \& Astrophysics, Enrico Fermi Institute, The University of Chicago, Chicago, IL 60637, USA}

\author{Marco Raveri}
\affiliation{Center for Particle Cosmology, Department of Physics and Astronomy, University of Pennsylvania, Philadelphia, PA 19104, USA}

\begin{abstract}
The canonical acoustic dark energy model (cADE), which is based on a scalar field with a canonical kinetic term that rapidly converts potential to kinetic energy around matter radiation equality, alleviates the Hubble tension found in $\Lambda$CDM. 
We show that it successfully passes new consistency tests in the CMB damping tail provided by the ACT data, while being increasingly constrained and distinguished from alternate mechanisms by the improved CMB acoustic polarization data from Planck. 
The best fit cADE model to a suite of cosmological observations, including the SH0ES $H_0$ measurement, has $H_0=70.25$ compared with $68.23$ (km\,s$^{-1}$\,Mpc$^{-1}$) in $\Lambda$CDM and a finite cADE component is preferred at the $2.8\sigma$ level. 
The ability to raise $H_0$ is now mainly constrained by the improved Planck acoustic polarization data, which also plays a crucial role in distinguishing cADE from the wider class of early dark energy models.   
ACT and Planck TE polarization data are currently mildly discrepant in normalization and drive correspondingly different preferences in parameters. 
Improved constraints on intermediate scale polarization approaching the cosmic variance limit will be an incisive test of the acoustic dynamics of these models and their alternatives.
\end{abstract}

\maketitle
\section{Introduction}
While the $\Lambda$CDM model of cosmology is remarkably successful at explaining a wide range of cosmological observations, it currently fails to reconcile distance-redshift measurements when anchored at low redshift through the distance ladder and high redshift by cosmic microwave background (CMB) anisotropies.
Specifically under $\Lambda$CDM, the Planck 2018 measurement of $H_0=67.4 \pm 0.5$ (in units of km s$^{-1}$ Mpc$^{-1}$ from here on)~\cite{Aghanim:2018eyx} is in $4.4\sigma$ tension with the latest SH0ES estimate $H_0=74.03 \pm 1.42$~\cite{Riess:2019cxk}.

The significance of this discrepancy makes it unlikely to be a statistical fluctuation and hence requires an explanation.
A resolution of this tension may lie in unknown systematic effects in the local distance ladder and an array of alternative measurement methods are being pursued to address this possibility. For example, calibrations based on the tip of the red giant branch~\cite{Freedman:2019jwv,Freedman:2020dne}, Mira variables~\cite{Huang:2019yhh},  megamasers~\cite{Pesce:2020xfe}, lensing time delays~\cite{Wong:2019kwg,Birrer:2020tax}  all give broadly consistent results but differ in the significance of the discrepancy.

On the other hand, different CMB measurements from Planck~\cite{Aghanim:2018eyx}, the South Pole Telescope (SPT)~\cite{Aylor:2017haa} and the Atacama Cosmology Telescope (ACT)~\cite{Aiola:2020azj,Choi:2020ccd} give compatible distance calibrations under $\Lambda$CDM.
While Planck and previous measurements weight the calibration to mainly the sound horizon, recent ground based experiments such as SPT and ACT
provide precise measurements of the damping scale as well~\cite{Aiola:2020azj}.

Cosmological solutions now generally require a consistent change in the calibration of both the CMB sound horizon and damping scale as standard rulers which anchor the high redshift end of the distance scale, preventing high redshift solutions that substantially change their ratio~\cite{Wyman:2013lza,Dvorkin:2014lea,Leistedt:2014sia,Ade:2015rim,Lesgourgues:2015wza,Adhikari:2016bei,DiValentino:2016hlg,Canac:2016smv,Feng:2017nss,Oldengott:2017fhy,Lancaster:2017ksf,Kreisch:2019yzn,Park:2019ibn}.
Once anchored there, the rungs on the distance ladder through baryon acoustic oscillations (BAOs) to supernovae Type IA (SN) leave little room for missing cosmological physics in between (see e.g.~\cite{Bernal:2016gxb, Aylor:2018drw, Raveri:2019mxg, Knox:2019rjx, Benevento:2020fev,Zhao:2017cud,Lin:2018nxe,Benevento:2019,Keeley:2019esp,Li:2019yem,Ghosh:2019tab,Liu:2019awo,Li:2020ybr,Dai:2020rfo,Zumalacarregui:2020cjh,Ballesteros:2020sik,Alestas:2020mvb,Jedamzik:2020krr,Braglia:2020iik,Gonzalez:2020fdy,DiValentino:2020kha}).

For this reason a class of models which posit a new form of energy density whose relative contribution peaks near matter radiation equality~\cite{Poulin:2018cxd,Lin:2019qug} have received much interest~\cite{Agrawal:2019lmo,Alexander:2019rsc,Smith:2019ihp,Berghaus:2019cls,Niedermann:2019olb,Sakstein:2019fmf,Ye:2020btb,Braglia:2020bym,Niedermann:2020dwg,Ye:2020oix}.
In these models, adding extra energy density changes the expansion rate before recombination and so the sound horizon while simultaneously tuning the timing of these contributions adjusts the damping scale as well.
These models can therefore successfully raise $H_0$ by changing the distance ladder calibration and are limited mainly by the compensating changes to parameters in order to offset the driving of the acoustic oscillations from the Jeans-stable additional component.   

These changes cause testable effects on CMB polarization, for modes that cross the horizon near matter-radiation equality~\cite{Lin:2019qug},
and on the clustering of cosmological structure, changing the amplitude and shape of the power spectrum~\cite{Ivanov:2020ril,Hill:2020osr,DAmico:2020ods,Niedermann:2020qbw}.
Differences between models in this class can also be distinguished by these effects.

Given the recent improvements in their measurement, we focus on the CMB polarization effects here and their implications for the canonical acoustic dark energy (cADE) model~\cite{Lin:2019qug}, where a scalar field with a canonical kinetic term suddenly converts its potential energy to kinetic energy by being released from Hubble drag on a sufficiently steep potential. With only two additional parameters, this model provides the most efficient and generic realization of the extra energy density scenarios.

This paper is organized as follows.
In \S~\ref{Sec:ADErecap} we briefly review the cADE  model and its relationship to other models in the literature.
In \S~\ref{sec:data} we introduce the data sets that we use to obtain the constraints presented in \S~\ref{Sec:Results}.
We highlight the role of ACT in \S~\ref{Sec:MinusACT}, Planck polarization in \S~\ref{Sec:Planck}, SH0ES in \S~\ref{Sec:MinusH0} 
and discuss differences with  other models where extra dark energy alleviates the Hubble tension in \S~\ref{Sec:Models}.
We conclude in \S~\ref{Sec:Conclusions}.

\section{Acoustic Dark Energy} \label{Sec:ADErecap}
In this section we review the model parameterization of acoustic dark energy (ADE) and its relationship to 
early dark energy (EDE) \cite{Poulin:2018dzj} following Ref.~\cite{Lin:2019qug}.   
For the purposes of this work, acoustic dark energy can be viewed  either as a dark fluid component described by 
an equation of state $w_\ADE$  and  rest frame sound speed $c_s^2$~\cite{Hu:1998kj} that becomes transiently important around matter radiation equality or as a 
scalar field that suddenly converts its potential energy to kinetic energy by being released from Hubble
drag at that time.  Adopting  the former description,
we model the ADE equation of state as
\begin{equation} \label{eqn:eos}
1+w_{\ADE}(a) = \frac{1+\wn }{[1+(a_c/a)^{3(1+\wn )/p}]^{p}} \,,
\end{equation}
which defines its energy density 
\begin{equation}
\rho_{\ADE}(a) = \rho_{\ADE}(a=a_c) 
      e^{-3\int_{a_c}^a [1+w_{\ADE}(\tilde a)] d\ln \tilde a}
\end{equation} 
once normalized to its fractional energy density contribution  at $a_c$
\begin{equation}
f_c = \frac{\rho_\ADE(a_c)}{\rho_{\rm tot}(a_c)} \,.
\end{equation}
The ADE component therefore {has a transition in} 
its equation of state around  a scale factor $a=a_c$ from $w_{\ADE}=-1$ to
$\wn$ which causes its fractional energy density to peak near $a_c$.    The rapidity of the transition is determined by $p$, which we set $p=1/2$ throughout as its specific value does not affect our qualitative results \cite{Lin:2019qug}. 
The connection to the scalar field picture comes from these asymptotic behaviors.   Given a constant sound speed, $\wn=c_s^2$ for a potential to kinetic conversion.   We call the case of a canonical scalar field
where $c_s^2=1$ ``cADE". In \S \ref{Sec:Models}, we widen the description to allow $\wn$ and $c_s^2$ to be free parameters and call this superset ``ADE".  
In summary, cADE is described by two parameters $\{ a_c, f_c \}$ whereas ADE is described
by four parameters $\{ \wn,  a_c, f_c, c_s^2 \}$.  When varying these parameters we impose flat, range bound priors:
$-4.5\leq \log_{10}a_c \leq -3.0$, $0\leq f_c \leq 0.2$, $0\leq \wn \leq 3.6$ and $0\leq c_s^2 \leq 1.5$.

These ADE models can be contrasted with the EDE model in its fluid description \cite{Poulin:2018dzj}.   
In the EDE case, the fluid behavior is modeled on a scalar field that oscillates around the minimum of its potential whose
equation of state can likewise be parameterized by Eq.~({\ref{eqn:eos}).   In this case, $p=1$ and
 $\wn=(n-1)/(n+1)$ is a free
parameter associated with raising an axion or cosine like potential to the $n$th power, 
where  $\wn \approx 1/2$ was found to best relieve the Hubble tension \cite{Poulin:2018cxd}.   An additional parameter 
$\Theta_i$ models the initial position of the field in the potential and controls an effective, scale-dependent,
sound speed (see \cite{Poulin:2018dzj,Lin:2019qug}).    The EDE model is therefore parameterized by $\{ \wn, a_c, f_c, \Theta_i \}$.   
When varying these parameters we use the same priors as ADE for $\{ a_c , f_c\}$ but fix $\wn=1/2$ 
and impose a flat prior on $\Theta_i$ in its range $0 \le \Theta_i \le \pi$.

In addition to these  parameters, the full cosmological model that we fit to data also includes the six $\Lambda$CDM parameters: 
the angular size of the CMB sound horizon $\theta_{s}$, the cold dark matter density $\Omega_c h^2$, baryon density  $\Omega_b h^2$, the optical depth to reionization  $\tau$, the initial curvature spectrum normalization at $k=0.05$ Mpc$^{-1}$, $A_s$ and its tilt $n_s$.
All these parameters have the usual non-informative priors \cite{Aghanim:2019ame}. We fix the sum of neutrino masses to the minimal value (e.g.~\cite{Long:2017dru}).
We use the EDE and ADE implementation in  CAMB~\cite{Lewis:1999bs}
 and CosmoMC~\cite{Lewis:2002ah} codes, following~\citep{Lin:2019qug}.
We sample the posterior parameter distribution until the Gelman-Rubin convergence statistic~\cite{gelman1992} satisfies $R-1<0.01$ or better unless otherwise stated. 

\section{Datasets} \label{sec:data}
In this paper, we combine several data sets relevant to the  Hubble tension.  
We use the publicly available Planck 2018 likelihoods for the CMB temperature and polarization power spectra at small (Planck 18 TTTEEE) and large angular scales (lowl+lowE) and the CMB lensing potential power spectrum in the multipole range $40 \leq \ell \leq 400$~\citep{Aghanim:2018eyx, Aghanim:2019ame, Aghanim:2018oex}.
We then compare the results to the 2015 version of the same data set~\citep{Ade:2015xua,Aghanim:2015xee,Ade:2015zua} and examine the impact of the improved high-$\ell$  polarization data, which we sometimes refer to as ``acoustic polarization" to distinguish it from the low-$\ell$ reionization signature.

We combine Planck data with ACT data which measures  CMB temperature and polarization spectra out to higher multipoles~\citep{Aiola:2020azj}. 
We exclude the lowest temperature multipoles for ACT that would otherwise be correlated with Planck, following~\citep{Aiola:2020azj}.

To expose the Hubble tension, we consider the SH0ES measurement of the Hubble constant, $H_0=74.03\pm1.42$ (in units of km\,s$^{-1}$\,Mpc$^{-1}$ here and throughout)~\citep{Riess:2019cxk}.
To these data sets we add BAO measurements from BOSS DR12~\cite{Alam:2016hwk}, SDSS Main Galaxy Sample~\cite{Ross:2014qpa} and 6dFGS~\cite{Beutler:2011hx} and the Pantheon Supernovae (SN) sample~\cite{Scolnic:2017caz}.
These data sets prevent resolving the Hubble tension by modifying the dark sector only between recombination and the very low redshift universe~\cite{Aylor:2018drw}.  

Our baseline configuration which we call ``All" contains: CMB temperature, polarization and lensing, BAO, SN and $H_0$ measurements.    We then proceed to examine the impact of key pieces of this combination by
removing or replacing various data sets.   Specifically we consider the following cases:
\begin{itemize}
\item All = Planck+ACT+SH0ES+BAO+Pantheon
\item -ACT = All$-$ACT
\item -P18Pol = All, but  Planck 18 TTTEEE $\rightarrow$ Planck 18 TT
\item -H0 = All$-$SH0ES
\item P18$\rightarrow$15,-ACT = -ACT, but  Planck 2018$\rightarrow$2015. This is the default combination used in \cite{Lin:2019qug}.
\end{itemize}
When highlighting the impact of a specific  data  component $i$ below, we quote 
$\Delta\chi^2_i \equiv -2\Delta \ln {\cal L}_i$, relative to the appropriate maximum  total likelihood  (${\cal L}$) model under $\Lambda$CDM.   For example $\Delta\chi^2_{\rm P}$ denotes the contribution from Planck CMB power spectra and includes  Planck TTTEEE+lowl+lowE, except for  the -P18Pol configuration where it includes Planck TT+lowl+lowE.
Since the prior on the additional ADE parameters is not physically motivated we do not consider evidence based comparison of model performances.

\section{Results} \label{Sec:Results}
In this section we discuss all results.   
In \S~\ref{Sec:All}, we present results for cADE and the All data combination.  
In \S~\ref{Sec:MinusACT} and \S~\ref{Sec:Planck} we explore the impact of the ACT and 2018 improvements to the Planck data, highlighting the crucial role of polarization.  
In \S~\ref{Sec:MinusH0}, we show that the ability to raise $H_0$ in cADE is not exclusively driven by the SH0ES measurement.   
We discuss how polarization measurements distinguish between cADE and the wider class of ADE and EDE models in \S~\ref{Sec:Models}.

\begin{table*}
\centering
\begin{tabular}{c|ccccc}
\hline
cADE  & All & -ACT & -P18Pol & -H0 & P$18\rightarrow15$,-ACT \\
\hline
$f_c$                       & 0.072(0.068$^{+0.025}_{-0.022}$) & 0.081(0.070$^{+0.027}_{-0.024}$) & 0.105(0.110$\pm$0.030) & 0.050(0.027$^{+0.008}_{-0.027}$) & 0.086(0.082$^{+0.026}_{-0.023}$) \\
$\log_{10}a_c$              & -3.42(-3.43$^{+0.05}_{-0.07}$)  & -3.50(-3.50$^{+0.07}_{-0.06}$)  & -3.41(-3.39$^{+0.03}_{-0.10}$)  &  -3.42(-3.47$^{+0.24}_{-0.11}$)  & -3.45(-3.46$^{+0.05}_{-0.06}$) \\
\hline
$H_0$                       & 70.25(70.14$\pm$0.82)  & 70.60(70.19$\pm$0.86)  & 71.38(71.54$\pm$1.07)  & 69.19(68.50$^{+0.55}_{-0.93}$)  & 70.57(70.60$\pm$0.85) \\
$S_8$    & 0.841(0.839$\pm$0.013) & 0.841(0.839$\pm$0.013) & 0.846(0.845$^{+0.018}_{-0.015}$) & 0.842(0.833$^{+0.011}_{-0.012}$) & 0.843(0.842$\pm$0.013) \\
\hline
$\Delta\chi^2_{\rm P}$   & -0.2   & -1.5   & -4.3   & -1.7  & -4.7   \\
$\Delta\chi^2_{\rm ACT}$ & -1.8    &  --    & -4.3   & -1.0  &  --     \\
$\Delta\chi^2_{\rm tot}$    & -11.5  & -10.7  & -19.4  & -1.6  & -12.7  \\
\hline
\hline
$H_0^{{\rm \Lambda CDM}}$    & 68.23(68.17$\pm$0.38)  & 68.29(68.22$\pm$0.40) & 68.30(68.32$\pm$0.42)  & 67.80(67.73$\pm$0.39)  & 68.58(68.35$\pm$0.42) \\
$S_8^{{\rm \Lambda CDM}}$ & 0.815(0.818$\pm$0.010) & 0.812(0.814$\pm$0.010) & 0.814(0.813$\pm$0.011) & 0.826(0.827$\pm$0.010) & 0.819(0.819$\pm$0.010)\\
\hline
\end{tabular}
\caption{\label{T:cADE}
Maximum likelihood (ML) parameters and constraints (mean and the 68\% C.L. lower and upper limits) for the cADE model with different data sets.
$\Delta\chi^2$ values for ML cADE model are  quoted relative to the ML $\Lambda$CDM model for the same data set.  $\Delta\chi^2_{P}$
reflects the contribution of the Planck CMB datasets involved in each case: for the -P18Pol case this includes
the TT, lowl, and lowE likelihoods while for  P$18\rightarrow15$ this employs the Planck 15 versions of all likelihoods (see~\ref{sec:data}). 
For comparison, the $H_0$ and $S_8 \equiv \sigma_8(\Omega_m/0.3)^{1/2}$ values for $\Lambda$CDM model are also presented. 
}
\end{table*}

\begin{figure}
\centering
\includegraphics[width=\columnwidth]{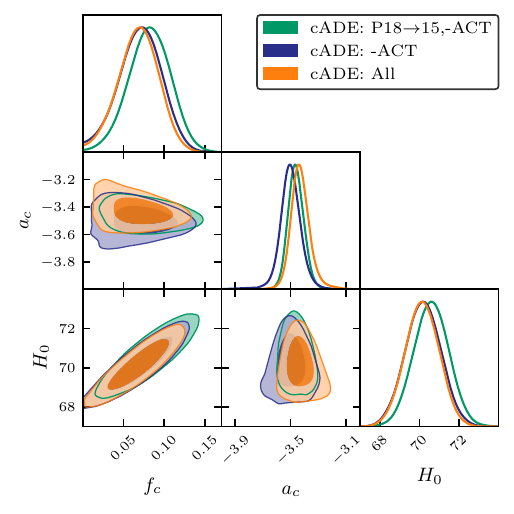}
\caption{
\label{fig:cADE-triangle1}
The marginalized joint posterior of parameters of the cADE model for  data sets that highlight the impact of ACT and the 2018 update to the Planck data. 
The darker and lighter shades correspond respectively to the 68\% C.L. and the 95\% C.L.
}
\end{figure}

\begin{figure}
\centering
\includegraphics[width=\columnwidth]{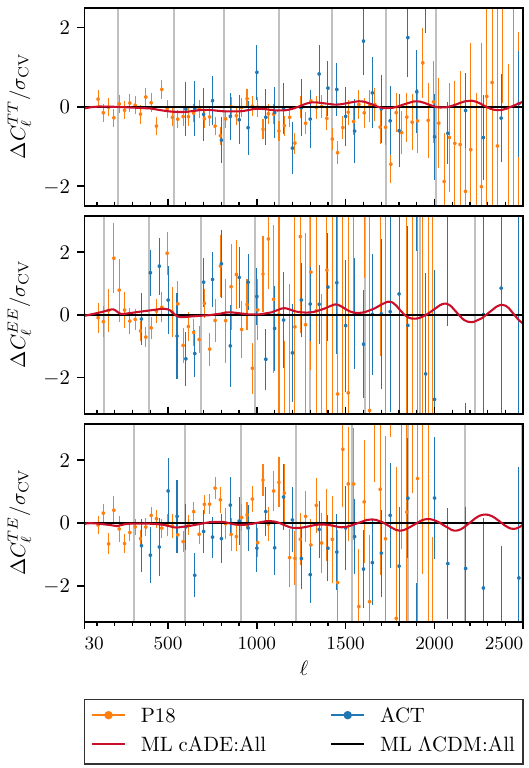}
\caption{\label{fig:P18+ACTresiduals+ACT}
CMB residuals of  Planck 2018 (orange points) and ACT (blue points) data and the ML cADE model (red curve) with respect to the ML $\Lambda$CDM model (black line) with both optimized to  All data.
The gray vertical lines indicate the positions of the acoustic peaks in the ML $\Lambda$CDM:All model. 
}
\end{figure}

\subsection{All data} \label{Sec:All}
We begin with results for the All data combination and the cADE model.   
In Fig.~\ref{fig:cADE-triangle1}, we show the constraints on the additional cADE parameters $f_c$ and $a_c$ as well as their impact on $H_0$.  
The mean value for $f_c$ is 2.8 standard deviations from zero, which we will  refer to as a $2.8\sigma$ detection,  and its distribution is strongly correlated with that of $H_0$.   
In Tab.~\ref{T:cADE} we also show the maximum likelihood (ML) parameters, notably $H_0=70.25$ in cADE vs.~68.23 in $\Lambda$CDM, as well as the improvement of fit over $\Lambda$CDM, a total of $\Delta \chi^2_{\rm tot} =- 11.5$ for 2 additional parameters.   
The portion that comes from Planck CMB power spectra, $\Delta \chi^2_{\rm P} = -0.2$ and from ACT $\Delta \chi^2_{\rm ACT}=-1.8$, reflects a slightly better fit to CMB power spectra than $\Lambda$CDM.  
Note that the ML value for $a_c$ is near matter-radiation equality.   Since the ML of a class of models
depends on the dataset it is optimized to, from this point forward we refer to such models as e.g.
ML cADE:All and ML $\Lambda$CDM:All.   

In Fig.~\ref{fig:P18+ACTresiduals+ACT}, we show the model and data residuals, both for Planck and ACT, of the ML cADE:All model relative to the  ML  $\Lambda$CDM:All model.   
The residuals are shown in units of $\sigma_{\rm CV}$, the cosmic variance error per multipole moment for the ML $\Lambda$CDM:All model
\begin{equation}
\sigma_{\rm CV} =
\begin{cases}
\sqrt{\frac{2}{2\ell+1}} C_\ell^{TT}, & {\rm TT} \,;\\
\sqrt{\frac{1}{2\ell+1}}  \sqrt{ C_\ell^{TT} C_\ell^{EE} + (C_\ell^{TE})^2}, & {\rm TE} \,;\\
\sqrt{\frac{2}{2\ell+1}}  C_\ell^{EE}, & {\rm EE} \,. \\
\end{cases}
\end{equation}
In spite of the higher $H_0$, the ML cADE:All model closely matches the ML $\Lambda$CDM:All model for all spectra and relevant multipoles.   Along the $f_c-H_0$ degeneracy,  $f_c$ adjusts the CMB sound horizon scale to match the acoustic peak positions while  $a_c$  near matter-radiation equality allows the damping tail of the CMB  to match as well.  
Note that the ACT data provide a new test, that has been successfully passed, for  this class of solution by providing more sensitive polarization constraints than Planck in the damping tail $\ell \gtrsim 10^3$ as we shall discuss in the next section.
Since adding an extra Jeans-stable energy density component drives CMB acoustic oscillations and changes the heights of the peaks, small variations in $\Omega_c h^2, \Omega_b h^2, n_s$ are required as well,
and these correlated changes remain mainly the same as those shown in Ref.~\cite{Lin:2019qug}}.
We shall see below that a crucial test that distinguishes cADE and related explanations of the Hubble tension is the imperfect compensation in the polarization, especially at intermediate multipoles that correspond to modes that cross the horizon near $a_c$ \cite{Lin:2019qug}.
Relatedly, as shown in Tab.~\ref{T:cADE}, the higher $\Omega_c h^2$ and $H_0$ values exacerbate the high $S_8=\sigma_8 (\Omega_m/0.3)^{1/2}$ values of $\Lambda$CDM so that accurate measurements of local structure test these scenarios as well~\cite{Ivanov:2020ril,Hill:2020osr,DAmico:2020ods,Niedermann:2020qbw}.

\begin{figure}
\centering
\includegraphics[width=\columnwidth]{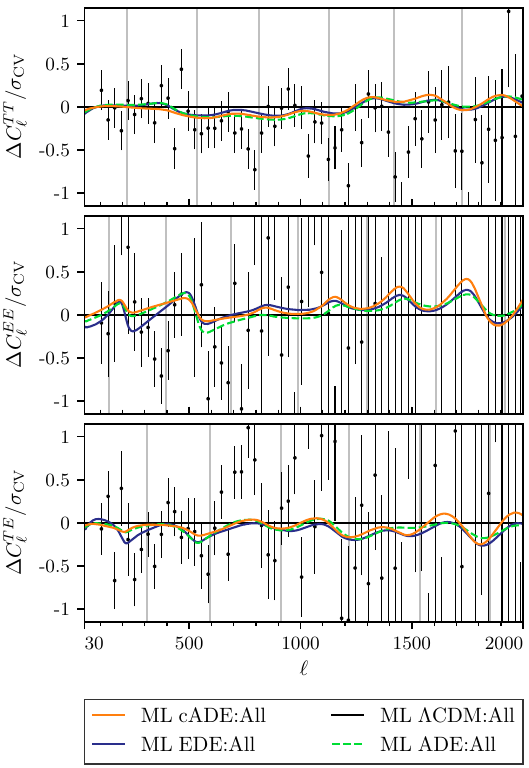}
\caption{\label{fig:CMBresiduals+ACT}
Planck CMB data residuals  and the  ML  cADE, ADE and EDE models relative to the ML  $\Lambda$CDM model, all optimized to All data.   
The gray vertical lines indicate the positions of the acoustic peaks in the ML $\Lambda$CDM:All model.
}
\end{figure}

\begin{figure}
\centering
\includegraphics[width=\columnwidth]{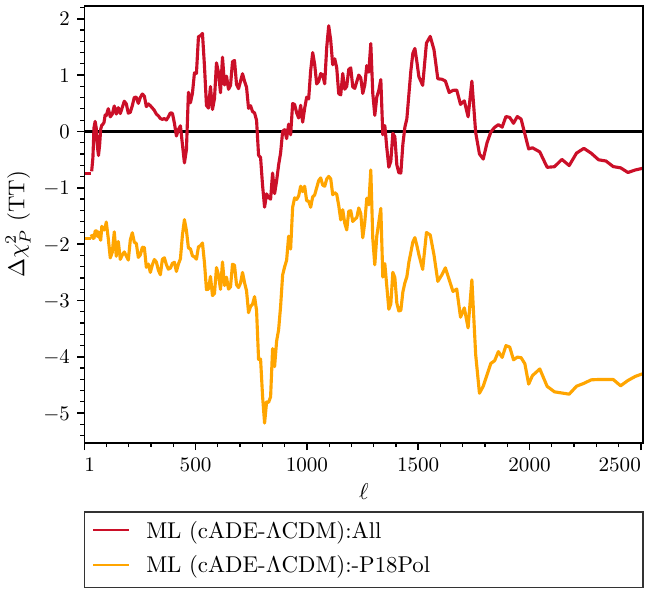}
\caption{\label{fig:chi2cum_TT_ALL+-POL}
Cumulative $\Delta\chi^2_{\rm P}$ for the Planck 18  TT+lowl+lowE likelihoods
 between ML cADE and ML $\Lambda$CDM models, optimized to either All or -P18Pol data.   Note that All includes the Planck 18 TTTEEE likelihood whereas -P18Pol does not.
}
\end{figure}

\subsection{ACT impact} \label{Sec:MinusACT}
The ACT data provide better constraints than Planck on the CMB EE polarization spectrum at $\ell \gtrsim 10^3$ as well as competitive TE  and corroborating TT constraints in this range.  The former provides new tests of the cADE model as shown in 
Fig.~\ref{fig:P18+ACTresiduals+ACT}.  On the data side, it is notable that for TT the  Planck data residuals compared with $\Lambda$CDM that oscillate with the acoustic peaks (gray lines) at $\ell \gtrsim 10^3$ are echoed in 
ACT data, albeit at a lower significance.   
We shall see below that were it not for Planck polarization constraints at lower multipole, the cADE fit to these oscillatory TT residuals would drive $H_0$ even higher.   
The additional constraining power of ACT polarization at high $\ell$ reduces the model freedom there and slightly shifts the compensation in acoustic driving toward higher $a_c$ and lower $\Omega_b h^2$.  This change in $a_c$
can be seen
in Fig.~\ref{fig:cADE-triangle1} where we also show the impact of removing ACT data.  
On the other hand the ability to raise $H_0$ is nearly unchanged.  

Interestingly, the ACT TE data is not in good agreement with the Planck data as noted in \cite{Aiola:2020azj} and attributed to $\sim 5\%$ calibration difference leading to $\Lambda$CDM parameter discrepancies at the  $2.7\sigma$ level. 
In Fig.~\ref{fig:P18+ACTresiduals+ACT}, we see that the Planck TE data have  residuals that oscillate with the acoustic
frequency  when compared with $\Lambda$CDM, whereas the ACT TE data do not.   
The ML cADE:All model attempts to compensate but must then compromise on the fit to the high-$\ell$ power spectra.   This tradeoff has
important implications for the comparison of cADE and $\Lambda$CDM as well as cADE and alternate models that add extra energy density near matter-radiation equality. 
This data discrepancy also motivates the study of the impact of Planck polarization data below.

\subsection{Planck impact} \label{Sec:Planck}
\subsubsection{Planck 2015 vs.~2018 data}
We start with the impact of the Planck 2018 data relative to the older 2015 release studied in \cite{Lin:2019qug} by reverting the data and removing ACT data in  P18$\rightarrow$15,-ACT.   
The main difference in the updated Planck data is the better polarization data and control over systematics, which makes both the TE and EE data important tests of the cADE model.

In Fig.~\ref{fig:cADE-triangle1} and Tab.~\ref{T:cADE}, we see that the main impact on cADE is a slight reduction of its ability to raise $H_0$ and a shift to lower $a_c$ that is countered by the ACT data in the All combination.   
This mild tension reflects the competition between fitting the high multipole spectra of both Planck and ACT and the intermediate multipole ($\ell \sim 500$) range of the Planck TEEE data.
The latter is a critical test of the cADE scenario since the perturbation scales associated with them cross the horizon near matter-radiation equality and are highly sensitive to changes in the manner the acoustic oscillations are driven.    
Polarization data represent a cleaner test than temperature  data since they lack the smoothing effects of the Doppler and integrated Sachs-Wolfe contributions.  
On the other hand, as we have seen Planck and ACT disagree somewhat on the TE spectrum in this range.  

In Fig.~\ref{fig:CMBresiduals+ACT} we highlight the Planck 2018 data residuals and ML cADE:All model residuals, both  relative to ML $\Lambda$CDM:All model.  Notice again the oscillatory residuals in TE and the features in cADE that respond to these residuals as well as the features in EE at $\ell \lesssim 600$.  
 
Furthermore, because of the ability to adjust Planck foregrounds, the overall amplitude of the TT data residuals, which have foregrounds fixed to the best fit to Planck 18 alone for visualization in Fig.~\ref{fig:CMBresiduals+ACT}, are low compared with the models.    
To better isolate the regions of the data that impact the models the most, we also show the cumulative $\Delta \chi^2_{\rm P}$ contributed by the Planck TT+lowl+lowE data 
in Fig.~\ref{fig:chi2cum_TT_ALL+-POL} for the ML cADE:All relative to ML $\Lambda$CDM:All model.
While the ML cADE:All model successfully minimizes differences with $\Lambda$CDM, there are  notable regions where the $\Delta\chi^2_P$ changes rapidly: $\ell \sim 500$, $800$, $1400$.    Note that the latter two regions are near the 3rd and 5th TT acoustic peaks and are related to the oscillatory TT residuals.
We shall next see that these areas reflect the trade-off between fitting the high $\ell$ power spectra of Planck and ACT and the intermediate scale polarization spectra of Planck.  

\begin{figure}
\centering
\includegraphics[width=\columnwidth]{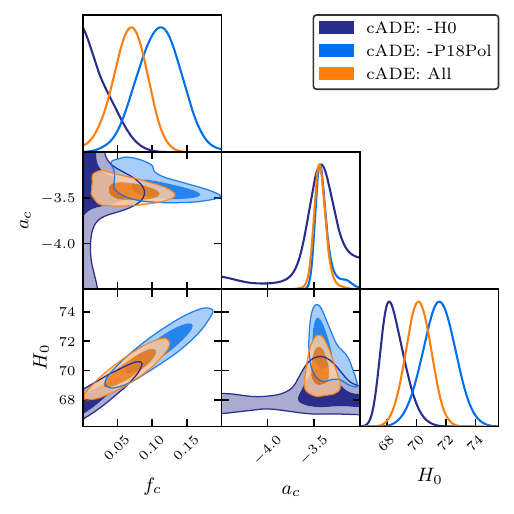}
\caption{\label{fig:show_22_cADE+ALL+-}
The marginalized joint posterior of parameters of the cADE model for  data sets that highlight the role of Planck 2018 acoustic polarization data and SH0ES $H_0$ measurements.  
The darker and lighter shades correspond respectively to the 68\% C.L. and the 95\% C.L.
}
\end{figure}

\begin{figure}
\centering
\includegraphics[width=\columnwidth]{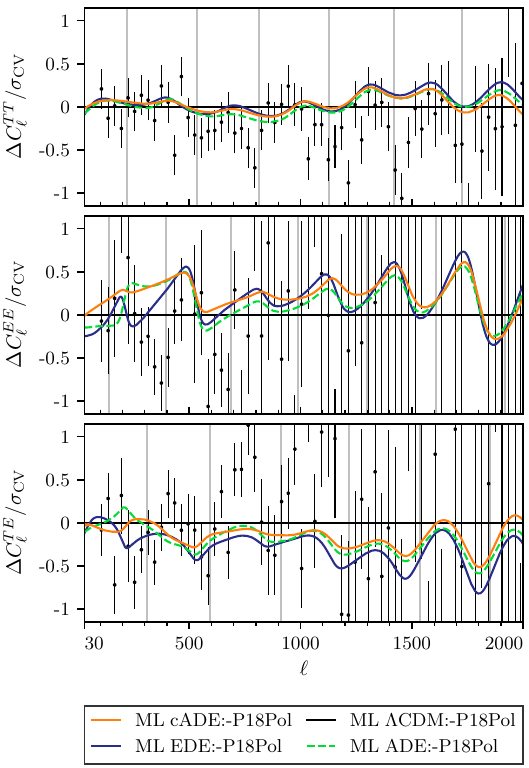}
\caption{\label{fig:CMBresiduals-POL+ACT}
Planck CMB data residuals and the ML cADE, ADE and EDE models relative to the ML $\Lambda$CDM model, all optimized to the -P18Pol data.   The gray vertical lines indicate the positions of the acoustic peaks in the ML $\Lambda$CDM:-P18Pol model.
}
\end{figure}

\subsubsection{ Planck polarization impact} \label{Sec:MinusPlanckPol}
The crucial role of intermediate scale TE and EE data in distinguishing models and the discrepancy between the TE calibrations of Planck and ACT motivate a more direct examination of the impact of Planck polarization data.    
In Fig.~\ref{fig:show_22_cADE+ALL+-} and Tab.~\ref{T:cADE},
we show the cADE parameters and $H_0$ constraints without the Planck 2018 acoustic polarization data but including acoustic polarization data from ACT as -P18Pol.
Notice that the ability to raise $H_0$ increases to $H_0=71.38$ for the ML cADE:-P18Pol model and total $\Delta \chi^2_{\rm tot}$ improvement  over $\Lambda$CDM rises to $-19.4$.  Correspondingly, a finite $f_c$ is preferred at $\sim 3.7\sigma$ and its ML value increases to $f_c=0.105$.
The fit to both the remaining Planck CMB power spectra data  and ACT temperature and polarization data correspondingly also improve by $-4.3$ and $-4.3$ respectively.    
The transition scale $a_c$ can also further increase in value, especially at lower $f_c$.

In Fig.~\ref{fig:CMBresiduals-POL+ACT}, we show how the ML cADE:-P18Pol 
model fits residuals in the Planck TT data relative to the ML $\Lambda$CDM:-P18Pol model.  Notice that the cADE model now responds to the oscillatory TT residuals.   
In Fig.~\ref{fig:chi2cum_TT_ALL+-POL}, we see that the main cumulative TT improvement comes from
$\ell \gtrsim 1400$.  

\begin{figure}
\centering
\includegraphics[width=\columnwidth]{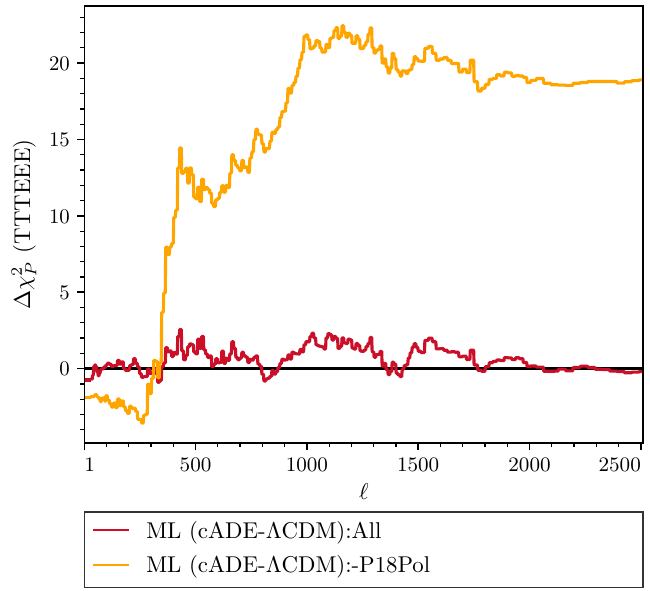}
\caption{\label{fig:chi2cum_TTTEEE_ALL+-POL}
Cumulative $\Delta\chi^2_{\rm P}$ for the Planck 18  TTTEEE +lowl+lowE likelihoods
 between ML cADE and ML $\Lambda$CDM models, optimized to either All or -P18Pol data.   
 Note that -P18Pol replaces the TTTEEE with the TT likelihood.
}
\end{figure}

On the other hand, Planck polarization data at intermediate scales $(\ell \sim 500)$ strongly disfavor this solution.
In Fig.~\ref{fig:chi2cum_TTTEEE_ALL+-POL}, we compare the cumulative Planck TTTEEE $\Delta\chi^2$ 
for the ML cADE:All model vs the ML cADE:-P18Pol model, both relative to their respective ML $\Lambda$CDM models.\footnote{The value of the cumulative $\Delta \chi^2_{\rm P}({\rm data})$ at the highest $\ell$ matches the values in  Tab.~\ref{T:cADE} only for the cases where the optimization matches the data, i.e.~ML cADE:-P18Pol in Fig.~\ref{fig:chi2cum_TT_ALL+-POL} and ML cADE:All in {Fig.~\ref{fig:chi2cum_TTTEEE_ALL+-POL}}.} 
While the former remains flat, reflecting an equally good fit for the cADE,
the latter encounters a sharp degradation in the fit just below $\ell \sim 500$ and a more gradual degradation between $500-1000$.  
The first degradation is associated with features in the EE spectrum and the second receives contributions from  the uniformly low TE spectrum
in Fig.~\ref{fig:CMBresiduals-POL+ACT}.
Since the Planck polarization data are far from cosmic variance limited even just statistically, future data in this region can provide a  sharp test of cADE and distinguish it from alternatives.

\subsection{SH0ES impact} \label{Sec:MinusH0}
Given the highly significant $H_0$ tension in $\Lambda$CDM, it is interesting to ask whether preference for
a higher $H_0$ in cADE simply reflects the SH0ES $H_0$ data.    In Fig.~\ref{fig:show_22_cADE+ALL+-} and Tab.~\ref{T:cADE}, we show the impact of removing this data.    Notice that although the ML value of $H_0$ drops to  69.19, the cADE constraints still allow a non-Gaussian tail to the higher $H_0$ values that are compatible with the All data.  The ML cADE:-H0 model remains a better fit to both Planck and ACT  temperature and polarization data than the ML $\Lambda$CDM:-H0 model which has a lower $H_0=67.8$.     
In this case, finite values for the cADE parameter $f_c$ are no longer significantly preferred.   Since all cADE models become indistinguishable from 
$\Lambda$CDM in the limit $f_c\rightarrow 0$, there is a large prior parameter volume associated with
the poorly constrained $a_c$ that favors $\Lambda$CDM, pulling the posterior probability of $H_0$ to lower
values and skewing the distribution.

\begin{table*}
\centering
\begin{tabular}{c|cccccc}
\hline
  & $\Delta\chi^2_{\rm tot}$ & $H_0$ & $f_c$ & $\log_{10}a_c$ & $\wn$ & $c_s^2$ or $\Theta_i/\pi$ \\
\hline
\fADE\  (ALL) & -14.0  & 70.25(69.67$^{+0.93}_{-0.97}$)  & 0.061(0.055$^{+0.028}_{-0.030}$)  & -3.60(-3.57$^{+0.20}_{-0.12}$) & 0.55(1.37$^{+0.37}_{-1.09}$) & 0.70(0.87$\pm$0.29) \\
EDE (ALL)  & -16.6  & 71.03(71.14$^{+0.98}_{-0.99}$)  & 0.056(0.061$^{+0.018}_{-0.017}$)  & -3.71(-3.68$^{+0.09}_{-0.07}$) & 0.5(fixed)  & 0.94($>$0.84) \\
\hline
\fADE\  (-ACT) & -11.9  & 70.55 & 0.074 & -3.61 & 0.68 & 0.80 \\
EDE (-ACT)  & -13.7  & 71.61 & 0.068 & -3.80 & 0.5(fixed)  & 0.92 \\
\hline
\fADE\  (-P18Pol) & -23.7  & 72.11  & 0.103 & -3.51 & 0.57 & 0.85 \\
EDE (-P18Pol)  & -26.1  & 73.07  & 0.100 & -3.65 & 0.5(fixed)  & 0.90 \\
\hline
\fADE\  (-H0) & -3.9  & 69.18  & 0.049 & -3.58 & 0.81 & 0.71 \\
EDE (-H0)  & -4.0  & 70.11  & 0.044 & -3.69 & 0.5(fixed) & 0.94 \\
\hline
\fADE\  (P18$\rightarrow$15,-ACT)   & -14.1   & 70.81(70.20$^{+0.87}_{-0.88}$) & 0.086(0.079$\pm$0.033) & -3.52(-3.50$^{+0.14}_{-0.08}$) & 0.87(1.89$^{+0.85}_{-1.07}$) & 0.86(1.07$^{+0.30}_{-0.20}$) \\
EDE (P18$\rightarrow$15,-ACT)    & -16.6   & 71.92(71.40$^{+1.07}_{-1.05}$) & 0.074(0.064$^{+0.020}_{-0.018}$) & -3.72(-3.72$^{+0.10}_{-0.07}$) & 0.5(fixed) & 0.90($>$0.82)\\ 
\hline
\end{tabular}
\caption{\label{T:ADEs}
ML parameters and constraints (mean and the 68\% C.L. lower and upper limits) for of cADE, \fADE, EDE models with different data sets. 
$\Delta\chi^2_{\rm tot}$ values are  quoted relative to the ML $\Lambda$CDM model for the same data set. 
 The column labeled ``$c_s^2$ or $\Theta_i/\pi$" indicates $c_s^2$ for \fADE\ and $\Theta_i/\pi$ for EDE. 
Since the boundary $\Theta_i/\pi=1$ is consistent with the data, we have quoted the 1-sided 68\% CL lower interval from this boundary.
Both \fADE\ and EDE have four parameters in addition to $\Lambda$CDM, but the $\wn$ value of EDE is crudely optimized by setting it to the value of best solving the $H_0$ tension following \cite{Poulin:2018dzj}.
}
\end{table*}
\subsection{Distinguishing model alternatives} \label{Sec:Models}
As we have seen in the previous sections, intermediate scale polarization data is crucial for 
limiting the ability of the cADE model to raise $H_0$ as well as distinguishing it from $\Lambda$CDM.  
This is because differences in acoustic driving are most manifest for modes that cross the horizon
while the additional energy density is important and the  signatures in polarization vs.~temperature spectra are clearer, due to the lack of other contaminating effects.

\begin{figure}
\centering
\includegraphics[width=\columnwidth]{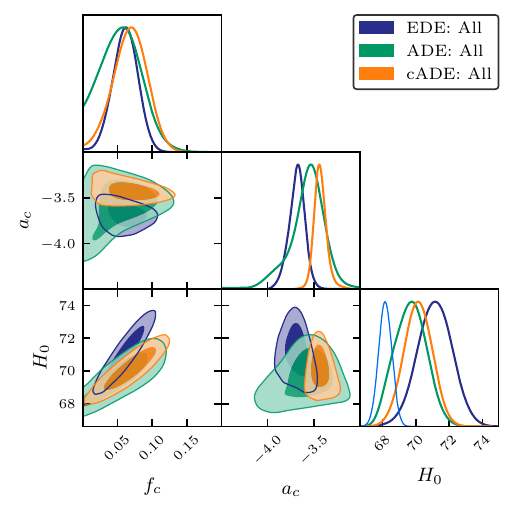}
\caption{\label{fig:show_22_ADE+default18+ACT}
The marginalized joint posterior of parameters of different models for the All data.
The darker and lighter shades correspond respectively to the 68\% C.L. and the 95\% C.L.
The 1-D $H_0$ distribution of $\Lambda$CDM: All chain is also shown as the light blue curve.
}
\end{figure}

Intermediate scale polarization is equally important for distinguishing cADE from the wider class of ADE models or EDE models.    
In Tab.~\ref{T:ADEs} we show results for these wider classes and the joint posterior of the common parameters along with $H_0$ are shown in Fig.~\ref{fig:show_22_ADE+default18+ACT}.   The ADE chains are converged at the $R-1<0.04$ level, reflecting degeneracies in poorly constrained parameters.
Note that both models possess 4 additional parameters, but for EDE we have followed \cite{Poulin:2018dzj} in crudely optimizing $\wn$ by setting it to $\wn =0.5$.   
In the ADE case the ML $H_0$ value remains at $H_0=70.25$ in cADE while in EDE case it rises to $71.03$.  
The total $\Delta \chi_{\rm tot}^2$ for the All dataset also improves from $-11.5$  to $-14.0$ and $-16.6$ respectively for 2 additional parameters.  
The All dataset therefore does not strongly favor either increase in model complexity.    
Note that because of the large parameter volume in \fADE\ near $f_c\rightarrow 0$, the posterior of $H_0$ in that case is strongly pulled by the prior to lower values than the ML value.   For EDE notice that in the scalar field interpretation
$\Theta_i/\pi=1$ is a field whose initial value is at the top of the potential and the data require a moderate tuning
to this boundary value \cite{Lin:2019qug}.

Tab.~\ref{T:ADEs}  also displays the  ML models for the various other data combinations discussed above.  
The trends are similar to those discussed for cADE.   In addition, for \fADE, the All data favors a lower
value for $\wn$ due mainly to the ACT data as compared with the P15-based previous results
from Ref.~\cite{Lin:2019qug}.  

More interestingly for the future, Figs.~\ref{fig:P18+ACTresiduals+ACT} and \ref{fig:CMBresiduals-POL+ACT} 
show that the current compromises between fitting the high $\ell$ power spectra of Planck and ACT vs.~the intermediate scale Planck polarization data are model dependent, especially in the polarization spectra
around $\ell \lesssim 500$.    
Since the Planck data are far from cosmic variance limited in TEEE, better measurements in this regime can distinguish between the various alternatives for adding extra energy density around matter radiation equality to alleviate the Hubble tension.

\section{Discussion} \label{Sec:Conclusions}
The acoustic dark energy model, which is based on a canonical kinetic term for a scalar field which rapidly converts potential to kinetic energy around matter radiation equality, alleviates the Hubble tension in $\Lambda$CDM 
and successfully passes new consistency tests in the CMB damping tail provided by the ACT data, while being increasingly constrained and distinguished from alternate mechanisms by the better intermediate scale polarization data from Planck.
The best fit cADE model has $H_0=70.25$ compared with $68.23$ in $\Lambda$CDM and a finite cADE component is preferred at the $2.8\sigma$ level.  
While this preference is driven by the SH0ES measurement of $H_0$ itself, even without this data the cADE model prefers a higher $H_0$ than in $\Lambda$CDM.

Intermediate scale $(\ell \sim 500)$ polarization data plays a critical role in testing these and other scenarios where an extra component of energy density alters the sound horizon and damping scale of the CMB.   
Such components also drive CMB acoustic oscillations leaving particularly clear imprints on the polarization of modes that cross the horizon around matter radiation equality.   
Were it not for the Planck 2018 polarization data, the ML cADE model would have $H_0=71.38$ and more fully resolve the Hubble tension.   
Intriguingly the ACT TE data do not  agree with Planck TE data in their normalization \cite{Aiola:2020azj}
and in cADE the two data sets drive moderately different preferences in parameters, especially
the epoch $a_c$ at which its relative energy density peaks.    
In the wider class of non-canonical acoustic dark energy (ADE) or early dark energy (EDE), which differ in the manner that acoustic oscillations are driven, polarization data at these scales is critical for distinguishing models, with the current freedom allowing an even larger $H_0 \sim 70-71$ and $71-73$ at ML with and without Planck polarization, albeit with two additional parameters.

Given the current statistical and systematic errors in measurements, future intermediate scale polarization data can provide even more incisive tests of the cADE model and its alternatives to resolving the Hubble tension. 

\acknowledgments
MXL and WH are supported by U.S.~Dept.~of Energy contract No. DE-FG02-13ER41958 and the Simons Foundation. 
MR is supported in part by NASA ATP Grant No. NNH17ZDA001N, and by funds provided by the Center for Particle Cosmology. 
Computing resources are provided by the University of Chicago Research Computing Center through the Kavli Institute for Cosmological Physics at the University of Chicago. 

\vfill

\bibliography{ADEpol}
\end{document}